\documentclass[11pt,nofootinbib,floatfix]{revtex4}
\usepackage{mathrsfs}
\usepackage{graphicx}
\usepackage{amsmath}
\usepackage{amsfonts}
\usepackage{amssymb}
\usepackage{color}

\usepackage{epsfig}
\usepackage{CJK}
\usepackage{eepic}
\usepackage{bbm}
\usepackage{dcolumn}% Align table columns on decimal point
\usepackage{bm}% bold math
\usepackage{ulem}
\usepackage{slashbox}
\usepackage{multirow}% for the tabular, added in 30.01.2013

\newcommand{\omits}[1]{}

\def\bc{\begin{center}}
\def\nno{\nonumber}
\def\ec{\end{center}}
\def\be{\begin{eqnarray}}
\def\ee{\end{eqnarray}}

\definecolor{dyellow}{rgb}{1.,0.8,.0}
\definecolor{myblue}{rgb}{.1,.1,.7}
\definecolor{dcyan}{rgb}{.0,.6,.6}
\definecolor{cyan}{rgb}{0.4,1.0,1.0}
\definecolor{dmagenta}{rgb}{0.6,0.0,0.6}
\definecolor{brown}{rgb}{0.6,0.2,0.}
\definecolor{darkblue}{rgb}{.0,.0,0.5}
\definecolor{darkred}{rgb}{0.75,0.0,0.0}
\definecolor{orange}{rgb}{1.,.6,.0}
\definecolor{dorange}{rgb}{0.8,.4,.0}
\definecolor{green}{rgb}{0.0,1.0,0.0}
\definecolor{darkgreen}{rgb}{0.0,0.6,0.0}
\definecolor{purple}{rgb}{.4,.0,.4}
\definecolor{lightgrey}{rgb}{0.7, 0.7, 0.7}
\definecolor{grey}{rgb}{0.4, 0.4, 0.4}
\def\Si{\Sigma}

\def\al{\alpha}

\def\la{\lambda}

\def\Na{\nabla}

\begin{document}

%\preprint{hep-th/yymmnnn}

\title{Holographic Interpretation of Acoustic Black Holes}

\author{Xian-Hui Ge$^{2}$} \email{gexh@shu.edu.cn}
\author{Jia-Rui Sun$^{1,6}$} \email{sunjiarui@sysu.edu.cn}
\author{Yu Tian$^{3,6}$} \email{ytian@ucas.ac.cn}
\author{Xiao-Ning Wu$^{4,6,7}$} \email{wuxn@amss.ac.cn}
\author{Yun-Long Zhang$^{5}$} \email{zhangyunlong@ntu.edu.tw}

\affiliation{${}^1$Institute of Astronomy and Space Science, Sun Yat-Sen University, Guangzhou 510275, China}
\affiliation{${}^2$Department of Physics, Shanghai University, Shanghai 200444, China}
\affiliation{${}^3$School of Physics, University of Chinese Academy of Sciences, Beijing 100049, China}
\affiliation{${}^4$Institute of Mathematics, Academy of Mathematics and System Science, Chinese Academy of Sciences, Beijing 100190, China}
\affiliation{${}^5$Department of Physics and Center of Advanced Study in Theoretical Sciences, National Taiwan University, Taipei 10617, Taiwan}
\affiliation{${}^6$State Key Laboratory of Theoretical Physics, Institute of Theoretical Physics, Chinese Academy of Science, Beijing 100190, China}
\affiliation{${}^7$Hua Loo-Keng Key Laboratory of Mathematics, CAS, Beijing 100190, China}

%\date{May, 2014}

%% REVTEX4
%\maketitle

\begin{abstract}
  With the attempt to find the holographic description of usual acoustic black holes in fluid, we construct an acoustic black hole formed in the $d$-dimensional fluid located at the timelike cutoff surface of a neutral black brane in asymptotically AdS$_{d+1}$ spacetime, the bulk gravitational dual of the acoustic black hole is presented at first order of the hydrodynamic fluctuation. Moreover, the Hawking-like temperature of the acoustic black hole horizon is showed to be connected to the Hawking temperature of the real AdS black brane in the bulk, and the duality between the phonon scattering in the acoustic black hole and the sound channel quasinormal mode propagating in the bulk perturbed AdS black brane is extracted. We thus point out that, the acoustic black hole appeared in fluid, which was originally proposed as an analogous model to simulate Hawking radiation of the real black hole, is not merely an analogy, it can indeed be used to describe specific properties of the real AdS black holes, in the spirits of the fluid/gravity duality.

\end{abstract}

%% REVTEX4
%\pacs{}

\maketitle
\newpage
\tableofcontents

%%%%%%%%%%%%%%%%%%%%%%%%%%%%%%%%%%%%%%%%%%%%%%%%%%%%%%%%%%%%%%%%%%%%%%
\section{Introduction}\label{sec:intro}
%%%%%%%%%%%%%%%%%%%%%%%%%%%%%%%%%%%%%%%%%%%%%%%%%%%%%%%%%%%%%%%%%%%%%%
Searching for the relationship between gravity and fluid has a long
history. The original study dates back to the the late 1970s, during which the black hole membrane paradigm was developed \cite{Damour:1978cg} (see also \cite{Thorne:1986iy,Price:1986yy}). It was showed that in the membrane paradigm formalism, the black hole can be regarded as an viscous fluid living on the null or timelike surface (membrane) on or outside its horizon, while the membrane actually acts as a boundary or a cut-off surface of the black hole spacetime with appropriate boundary conditions and it captures information which can be used as an effective description of the physics inside itself. Later on, Unruh showed that, for the nonrelativistic, irrotational inviscid moving fluid, the equation of motion governing the dynamics of the sound mode (phonons) can be expressed as a massless Klein-Gorden equation in an effective spacetime background containing the sonic horizon when the local fluid velocity exceeds the speed of sound, which resembles the real black hole. Consequently, a Hawking-like temperature can be defined for the sonic horizon analogous to the real black hole, so it was named as the acoustic black hole \cite{Unruh:1980cg}. However, although the acoustic black hole possesses many characteristics that resemble the real black hole system, it seems that its dynamics, governed by the Euler or the Navier-Stokes equation and equation of continuity (for relativistic fluid, the EoMs are conservation equations), has nothing to do with that of the latter, determined by the Einstein equation (plus dynamical equations of the background matter fields). Therefore, the acoustic black hole was merely regarded as an analogous gravitational model to mimic the phenomena in real gravitational systems, and the testing of Hawking-like radiation in acoustic black holes doesn't mean the detecting of the Hawking radiation from real black holes. Even though, topics on analogous gravity still received much attention during the past years, both theoretically and experimentally, with the attempt to obtain some insights for studying the real gravitational systems, for example, the emergence of acoustic black hole from the Bose-Einstein condensation \cite{Garay:1999sk,Barcelo:2001ca}, the superfluid helium-3 and other cold bosonic systems \cite{Volovik:2002ci}, and from the superconductors \cite{Ge:2010wx,Ge:2010eu}, e.g., see an nice up-to-date review \cite{Barcelo:2005fc} and references therein.

An interesting question is, can we add new interpretations to the word ``analogous'', e.g., can we really gain the information of a real black hole from an acoustic black hole made in the laboratory? The answer is probably yes. When taking the holographic principle into account, it is natural to expect that there exists a bulk holographic description of the acoustic black hole emerged from the fluid on the boundary in the context of the fluid/gravity correspondence~\cite{Policastro:2002se,Policastro:2002tn,Son:2007vk} (which is the low frequency and long wavelength version of the gauge/gravity duality~\cite{'tHooft:1993gx,Susskind:1994vu,Maldacena:1997re,Gubser:1998bc,Witten:1998qj}) . More specifically, we can ask what will happen in the bulk black hole when the supersonic phenomenon appears in its dual fluid system (or any finite temperature interacting field theory in the long time and long wavelength limit) on the boundary of the asymptotically AdS spacetime? According to the fluid/gravity correspondence, the fluid located at the
asymptotic boundary has the same temperature and entropy with those of its dual bulk black hole. Furthermore, it was showed that the Navier-Stokes equation of the boundary fluid is also dual to the long wavelength behavior of the Einstein equation of the bulk gravity \cite{Bhattacharyya:2008jc,Bredberg:2011jq}. Hence, when an acoustic black hole forms in the boundary fluid, besides matching of the fluid temperature and the entropy with those of its corresponding bulk black hole (note that the temperature of the acoustic horizon, determined by the gradient of the fluid velocity at the acoustic horizon, is different with the temperature of the fluid, while their relation is determined by the EoMs of the fluid), its dynamics can also in principle be fully determined from the dynamics of the bulk gravitational theory, and the linearized normal mode fluctuations of the boundary fluid (which are described by the sound mode, shear mode and the transverse traceless mode) correspond to the linearized perturbations of the dual bulk gravity and gauge fields (which are the scalar mode, vector mode and the tensor mode in the long time and long wavelength limit together with the quasinormal mode boundary condition), respectively \cite{Kovtun:2005ev}. Thus, for a compressible fluid on the boundary or cutoff surface of the AdS black brane, there should exist a phonon/scalar quasinormal mode correspondence. Now that the phonon also propagates into the acoustic black hole emerging from the fluid, it is expectable that the acoustic black hole can probably be related to or mapped to a real black hole in the asymptotically AdS spacetime. In this paper, we give the derivation to construct a $d$-dimensional acoustic black brane formed in the fluid located at the finite timelike cutoff surface in an neutral black brane in asymptotically AdS$_{d+1}$ spacetime \footnote{Preliminary attempt on the related problem see, \cite{Das:2010mk}, where an acoustic black hole in 4-dimensional conformal ideal fluid at the AdS boundary was analyzed.}, in the spirit of the Wilsonian approach to the fluid/gravity correspondence \cite{Bredberg:2010ky,Faulkner:2010jy,Brattan:2011my,Pinzani-Fokeeva:2014cka}. We show that, the acoustic black hole geometry can be obtained from the correspondence between the dynamics which govern the fluid and the gravity, namely, the equivalence between the conservation equation of fluid at cutoff surface and the constraint equations of the Einstein equation of bulk AdS black brane. Besides, the Hawking-like temperature of the acoustic black hole horizon can indeed be connected to the real Hawking temperature of its dual bulk AdS black brane, which may give strong supports to various studies on detecting the Hawking-like radiation from the acoustic black hole. Furthermore, we also find that the normal mode excitation in the acoustic black hole-the phonon, is dual to the sound channel of the quasinormal mode in the bulk AdS black brane with first order hydrodynamic fluctuations. Based on these results, the acoustic black hole can indeed be used to study certain bulk gravitational (plus the possible bulk matter fields) perturbations, i.e. the scalar quasinormal mode perturbation, and the appearance of the supersonic phenomenon might introduce testable effect in its dual bulk AdS black hole. In this sense, the acoustic black hole is no longer just an analogous model of the real black hole.

The rest parts of this paper is organized as follows. In section \ref{sec:NR acoustic}, we give a brief review of the original derivation of acoustic black holes from the general $d$-dimensional non-relativistic inviscid and viscous fluid, respectively. In section \ref{sec:Rela acoustic}, we obtain the acoustic black holes in the relativistic fluid with first order dissipations. Then in Section \ref{sec:holo acoustic}, we present a holographic realization of $d$-dimensional acoustic black hole formed in fluid at finite timelike cutoff surface in asymptotically AdS$_{d+1}$ spacetime, based on the fluid/gravity correspondence. Various properties of the holographic acoustic black hole are studied. Especially, we show its dual bulk gravitational geometry, determine the relation between temperature of the acoustic horizon and that of the dual real black hole horizon, and show there is a duality between the phonon in the acoustic balck hole and the scalar quasinormal mode of the bulk AdS black brane. The conclusions and discussions are drawn in Section \ref{sec:conclusion}. In the appendix A and B, we list the explicit form of the first order metric corrections to the bulk AdS black brane and a 3-dimensional holographic relativistic rotating acoustic black hole at the cutoff surface, respectively.

%%%%%%%%%%%%%%%%%%%%%%%%%%%%%%%%%%%%%%%%%%%%%%%%%%%%%%%%%%%%%%%%%%%%%
\section{Acoustic metric from non-relativistic fluid}\label{sec:NR acoustic}
%%%%%%%%%%%%%%%%%%%%%%%%%%%%%%%%%%%%%%%%%%%%%%%%%%%%%%%%%%%%%%%%%%%%%%
\subsection{Inviscid acoustic black brane}
%%%%%%%%%%%%%%%%%%%%%%%%%%%%%%%%%%%%%%%%%%%%%%%%%%%%%%%%%%%%%%%%%%%%%%
For the non-relativistic $d$-dimensional convergent locally vorticity free (irrotational) neutral and
inviscid fluid moving in the Minkowski spacetime, its dynamics is governed by the Euler equation and the equation of continuity
\be\label{nr-eoms} &&\rho\left(\partial_t
\vec{v}+\vec{v}\cdot\nabla\vec{v}\right)=-\nabla
p-\rho\nabla\Phi,\nno\\
&& \partial_t\rho+\nabla\cdot(\rho \vec{v})=0,\ee
where $\Phi$ is the potential field of the external force such as the Newtonian gravitational field. Since $\nabla\times \vec{v}=0$, the velocity can be described by the gradient of a potential field, i.e. $\vec{v}=\nabla\psi$. Considering small hydrodynamic fluctuations of the
background fluid up to linear order while keeping the external force potential $\Phi$ fixed \footnote{Note that this linearized perturbation is different with the derivative expansion of the hydrodynamic variables, which will not alter the configuration of the background fluid, namely, the background fluid is still the ideal fluid.}
\be\label{1stp} \xi=\bar{\xi}+\delta\xi \quad {\rm and}\quad \psi=\bar{\psi}+\delta\psi,\ee
where $\xi=\ln\rho$. Then the EoM of the sound wave (phonon) is just the
Klein-Gorden equation of a massless scalar field
\be
\frac{1}{\sqrt{-\bar{g}}}\partial_\mu\left(\sqrt{-\bar{g}}\bar{g}^{\mu\nu}\partial_\nu\delta\psi\right)=0\ee
propagating in an effective acoustic geometry background \cite{Unruh:1980cg}
\be\label{a1}
ds_{\rm ac}^2=\left(\frac{\bar{\rho}}{c_{\rm s}}\right)^{\frac{2}{d-2}}\bigg(-(c_{\rm s}^2-\vec{\bar{v}}^2)dt^2-2\bar{v}_i dx^i dt+dx^idx_i\bigg),\ee
where $\sqrt{-\bar{g}}=\bar{\rho}^{\frac{d}{d-2}}c_{\rm s}^{\frac{-2}{d-2}}$, $\vec{\bar{v}}^2=\bar{v}_i \bar{v}_j \delta^{ij}$, $c_{\rm s}$ is the speed of sound, and
\be
\bar{g}^{\mu\nu}=\frac{1}{\sqrt{-\bar{g}}}\left(
              \begin{array}{cc}
                -\frac{\bar{\rho}}{c_{\rm s}^2} & -\frac{\bar{\rho}}{c_{\rm s}^2}\bar{v}_i \\\\
                -\frac{\bar{\rho}}{c_{\rm s}^2}\bar{v}_j & \frac{\bar{\rho}}{c_{\rm s}^2}\left(c_{\rm s}^2\delta_{ij}-\bar{v}_i \bar{v}_j\right) \\
              \end{array}
            \right).
\ee
Furthermore, making the coordinate transformation
\be\label{ttau} t=\tau-\int\frac{\bar{v}_{i}}{c_{\rm s}^2-\vec{\bar{v}}^2}dx^i,\ee
where the vorticity free condition ensures that $d^2$ acting on both sides of eq.(\ref{ttau}) give the same results, i.e. the zeros. The effective background eq.(\ref{a1}) becomes to
\be\label{a2}
ds_{\rm ac}^2&=&\left(\frac{\bar{\rho}}{c_{\rm s}}\right)^{\frac{2}{d-2}}\bigg(-(c_{\rm s}^2-\vec{\bar{v}}^2)d\tau^2+\frac{\bar{v}_i \bar{v}_j}{c_{\rm s}^2-\vec{\bar{v}}^2}
dx^i dx^j+dx^idx_i\bigg)\nno\\
&=&\left(\frac{\bar{\rho}}{c_{\rm s}}\right)^{\frac{2}{d-2}}\left(-\frac{c_{\rm s}^2}{\gamma_{\rm s}^2}d\tau^2+P^s_{ij}
dx^i dx^j\right),\ee
where $\gamma_{\rm s}=1/\sqrt{1-\vec{\bar{v}}^2/c_{\rm s}^2}$, $\bar{u}_{i}=\gamma_{\rm s}
\bar{v}_{i}/c_{\rm s}$ and the projecting operator is
$P^s_{ij}=\delta_{ij}+\bar{u}_{i}\bar{u}_{j}$.
In the spatially flat case, we can choose the coordinates such that the fluid velocity only has the $z-$th component, i.e. $\bar{v}_i=\bar{v}_{z}$, where $z \equiv x^{d-1}$, then we have
\be\label{a3}
ds_{\rm ac}^2&=&\left(\frac{\bar{\rho}}{c_{\rm s}}\right)^{\frac{2}{d-2}}\left(-(c_{\rm s}^2-\bar{v}_z^2)d\tau^2+\frac{\bar{v}_{z}^2}{c_{\rm s}^2-\bar{v}_z^2}
dz^2+dx^i dx_i\right)\nno\\
&=&\left(\frac{\bar{\rho}}{c_{\rm s}}\right)^{\frac{2}{d-2}}\left(-(c_{\rm s}^2-\bar{v}_z^2)d\tau^2+\frac{c_{\rm s}^2}{c_{\rm s}^2-\bar{v}_z^2}
dz^2+dx^a dx_a\right),\ee
where the spatial index $a$ runs from 1 to $d-2$. The acoustic black
brane appears when the local fluid velocity $\bar{v}_{z}$ exceeds the
speed of sound $c_{\rm s}$, let us further require that the fluid velocity only
depends on the coordinate $z$ and it becomes equal to $c_{\rm s}$ at
$z=z_{\rm sh}$, in the linear order approximation, we can expand $\bar{v}_{z}$ as
\be \bar{v}_{z}=c_{\rm s}+\frac{\partial \bar{v}_{z}}{\partial
z}\bigg|_{z_{\rm sh}}(z-z_{\rm sh})\equiv c_{\rm s}-\kappa(z-z_{\rm sh})
,\ee
requiring $\kappa>0$, thus when $z\leq
z_{\rm sh}$, $\bar{v}_{z}\geq c_{\rm s}$, consequently, phonons cannot escape
from the region $z\leq z_{\rm sh}$ and $z_{\rm sh}$ is the location of the
sonic horizon (where we have chosen the fluid velocity to be along the
direction of $-z$). Then the near acoustic horizon geometry of the acoustic black brane is
\be\label{a4} ds_{\rm ac}^2&=&\left(\frac{\bar{\rho}}{c_{\rm s}}\right)^{\frac{2}{d-2}}\left(-2\kappa
c_{\rm s}(z-z_{\rm sh})d\tau^2+\frac{c_{\rm s}}{2\kappa(z-z_{\rm sh})}
dz^2+dx^a dx_a\right),\ee
%
%&=&\bar{\rho}\left(-2\alpha
%(z-z_{sh})d\tau^2+\frac{1}{2\alpha(z-z_{sh})}
%dz^2+dx^2+dy^2\right),\ee
%
which resembles the near horizon geometry of the real black hole and can be further written in the Rindler spacetime. The temperature of the acoustic black hole is just
\be T_{\rm sh}=\left|\frac{\kappa}{2\pi}\right|=\frac{1}{2\pi}\left|\frac{\partial \bar{v}_{z}}{\partial z}\right|_{z_{\rm sh}}.\ee
%

%%%%%%%%%%%%%%%%%%%%%%%%%%%%%%%%%%%%%%%%%%%%%%%%%%%%%%%%%%%%%%%%%%%%%%
\subsection{Viscous acoustic black brane}
%%%%%%%%%%%%%%%%%%%%%%%%%%%%%%%%%%%%%%%%%%%%%%%%%%%%%%%%%%%%%%%%%%%%%%
The acoustic black hole can also be generalized into the viscous fluid, see \cite{Visser:1997ux}, in which the 4-dimensional fluid with shear viscosity was discussed. Here we consider the general $d$-dimensional viscous fluid with both the shear and bulk viscosities. The dynamics of such fluid is described by the Navier-Stokes and the conservation equations
\be\label{nrNS} \rho\left(\partial_t
\vec{v}+\vec{v}\cdot\nabla\vec{v}\right)&=&-\nabla
p+\eta\nabla^2\vec{v}+\left(\frac{d-3}{d-1}\eta+\zeta\right)\nabla(\nabla\cdot\vec{v})-\rho\nabla\Phi,\nno\\
 \partial_t\rho+\nabla\cdot(\rho \vec{v})&=&0.\ee
where $\eta$ is the shear viscosity and $\zeta$ is the bulk
viscosity of the fluid.

Similar to the inviscid fluid case, the EoM of the sound mode can be derived from taking the linearized hydrodynamic perturbations eq.(\ref{1stp}) to the fluid, then
\be\label{eomsoundvis}\Box \delta\psi=\frac{1}{\sqrt{-\bar{g}}}\partial_\mu\left(\sqrt{-\bar{g}}\bar{g}^{\mu\nu}\partial_\nu\delta\psi\right)
=-\left(\frac{2d-4}{d-1}\eta+\zeta\right)\left(\frac{c_{\rm s}}{\bar{\rho}}\right)^{\frac{2}{d-2}}
\frac{1}{c_{\rm s}^2}(\partial_t+\vec{\bar{v}}\cdot\nabla)\nabla^2 \delta\psi,\ee
which is a modified Klein-Gorden equation with higher derivative corrections, where $\nabla^2=\eta^{ij}\partial_i\partial_j$. However, the effective acoustic spacetime in which the sound modes propagate is the same as eq.(\ref{a1}).

In the Eikonal approximation $\delta\psi=a(x)\exp(-i\omega t+i\vec{k}\cdot \vec{x})$ (in which $a(x)$ varies slowly with respect to $x$), the dispersion relation of the sound mode obtained from the viscous acoustic black brane is
\be
&&\omega^2-2\vec{k}\cdot \vec{\bar{v}} \omega+\left(c_{\rm s}^2 \vec{k}^2-(\vec{k}\cdot \vec{\bar{v}})^2\right)=i\left(\frac{2(d-2)}{d-1}\nu+\mu\right)\left(\vec{k}\cdot \vec{\bar{v}}-\omega\right)\vec{k}^2 \\\label{frequencyac}
\Rightarrow \omega &=&\vec{k}\cdot \vec{\bar{v}} \pm \sqrt{c_{\rm s}^2 \vec{k}^2-\left(\frac{(d-2)}{d-1}\nu+\frac{\mu}{2}\right)^2(\vec{k}^2)^2}-\frac i 2 \left(\frac{2(d-2)}{d-1}\nu+\mu\right)\vec{k}^2\nno\\
&=&\vec{k}\cdot \vec{\bar{v}} \pm c_{\rm s} k-\frac{i}{2}  \left(\frac{2(d-2)}{d-1}\nu+\mu\right)\vec{k}^2 \mp \frac{1}{2c_{\rm s}}\left(\frac{(d-2)}{d-1}\nu
+\frac{\mu}{2}\right)^2\vec{k}^2k+\mathcal{O}(k^4),\ee
where $k=|\sqrt{\vec{k}^2}|$, and $\nu=\frac{\eta}{\bar{\rho}}$ and $\mu=\frac{\zeta}{\bar{\rho}}$ are the kinematic viscosities.

%%%%%%%%%%%%%%%%%%%%%%%%%%%%%%%%%%%%%%%%%%%%%%%%%%%%%%%%%%%%%%%%%%%%%%
\section{Acoustic black brane from relativistic fluid}\label{sec:Rela acoustic}
%%%%%%%%%%%%%%%%%%%%%%%%%%%%%%%%%%%%%%%%%%%%%%%%%%%%%%%%%%%%%%%%%%%%%%
The above acoustic black hole description of non-relativistic fluid can be accordingly generalized into the relativistic hydrodynamics, see e.g., the cases for ideal fluid in \cite{moncrief:1979,Bilic:1999sq}. We will extend the discussion into relativistic fluid with dissipations. For the $d-$dimensional neutral relativistic viscous fluid flowing in a curved spacetime, its stress tensor (in the first order expansion of the temperature and velocity fields) is
\be T^{\mu\nu}%&=&\epsilon u^\mu u^\nu +pP^{\mu\nu}-P^{\mu\alpha}P^{\nu\beta}\left(\eta(\nabla_\alpha u_\beta+\nabla_\beta u_\alpha)+\left(\zeta-\frac{2}{d-1}\eta\right)g_{\al\beta}\nabla^\la u_\la\right)\nno\\
=\epsilon u^\mu u^\nu +pP^{\mu\nu}-2\eta\sigma^{\mu\nu}-\theta\zeta P^{\mu\nu},\ee
where
\be \sigma^{\mu\nu}=P^{\mu\alpha}P^{\nu\beta}\left(\nabla_{(\alpha}u_{\beta)}-\frac{\theta}{d-1}P_{\alpha\beta}\right)
\quad {\rm and} \quad \theta=\nabla_\lambda u^\lambda
\ee
are the shear tensor and expansion associated with the velocity fields $u^\alpha$ and $P^{\mu\nu}=u^\mu u^\nu+g^{\mu\nu}$. The EoMs are obtained by projecting the conservation equation $\nabla_\mu T^{\mu\nu}=0$ along the longitudinal direction (equation of continuity)
\be\label{rela-continuity eq} \nabla_\mu(\epsilon u^\mu)+p\nabla_\mu u^\mu +\eta\bigg((u^\al \nabla_\al u^\nu) (u^\beta\nabla_\beta u_\nu)+(\nabla_\mu u^\al)(\nabla_\al u^{\mu})-u_\nu\nabla_\mu\nabla^\mu u^\nu\bigg)&&\nno\\
+\left(\zeta-\frac{2}{d-1}\eta\right)(\nabla_\mu u^\mu)^2=0&&\ee
and in the transverse direction (dynamical equation)
\be\label{rela-dyn eq} &&(\epsilon+p)u^\mu \nabla_\mu u_\la+P^\mu_{\;\la}\nabla_\mu p-\eta\bigg((\nabla_\mu u^\mu)(u^\alpha\nabla_\al u_\la)-(u^\al \nabla_\al u^\nu) (u^\beta\nabla_\beta u_\nu)u_\la+(\nabla_\mu u_\la)(u^\al\nabla_\al u^\mu)\nno\\
&&+(u^\mu\nabla_\mu u^\al)(\nabla_\al u_\la)+u^\mu u^\al\nabla_\mu\nabla_\al u_\la+\Na_\mu\Na^\mu u_\la+\Na_\mu\Na_\la u^\mu+ (u_\nu\Na_\mu\Na^\mu u^\nu)u_\la+(u_\nu\Na_\mu\Na^\nu u^\mu)u_\la \bigg)\nno\\
&&-\left(\zeta-\frac{2}{d-1}\eta\right)\bigg(\Na_\la\Na_\al u^\al+(u^\mu\Na_\mu u_\la)(\Na_\al u^\al)+ (u^\nu\Na_\nu\Na_\al u^\al)u_\la \bigg)=0.\ee
It is easy to see that the non-relativistic limit, namely, $u_\al=(1,\vec{v})$, $|\vec{v}|\ll 1$, $p\ll \epsilon$, $\vec{v}\frac{dp}{dt}\ll \Na p$ and $\epsilon=\rho +\frac 1 2 \rho \vec{v}^2+\rho \varepsilon \rightarrow \rho $ ($\varepsilon$ is the internal energy density and we have set the speed of light $c=1$), together with taking the flat spacetime limit $g_{\mu\nu}\rightarrow \eta_{\mu\nu}$, eqs.(\ref{rela-continuity eq})(\ref{rela-dyn eq}) reduce to the Navier-Stokes equation and the equation of continuity eq.(\ref{nrNS}) of the non-relativistic fluid.

For the conformal fluid flowing in the conformally flat spacetime we have
\be
\zeta=0,\quad p=\frac{\epsilon}{d-1},\quad \epsilon=\sigma T^d\quad {\rm and}\quad c_{\rm s}^2=\frac{1}{d-1},\ee
the corresponding acoustic metric is the same as that for ideal fluid (up to an numerical conformal factor) \cite{moncrief:1979,Bilic:1999sq}
\be\label{acour}ds_{\rm ac}^2=\left(\frac{\bar{T}^{d-2}}{c_{\rm s}}\right)^{\frac{2}{d-2}}\left(-c_{\rm s}^2 \bar{u}_\mu \bar{u}_\nu + P_{\mu\nu}\right)dx^\mu dx^\nu.\ee
Like the situation in the non-relativistic fluid, the presence of viscosities will not alter the acoustic geometry eq.(\ref{acour}). Instead, they will break the Lorentz symmetry of the fluid. The corresponding EoM for the phonon is
\be\label{eomsoundvisre}\Box \delta\psi=\frac{1}{\sqrt{-\bar{g}}}\partial_\mu\left(\sqrt{-\bar{g}}\bar{g}^{\mu\nu}\partial_\nu\delta\psi\right)
=-\left(\frac{2d-4}{d-1}\right)\frac{\eta}{\bar{T} \bar{s}}\left(\frac{c_{\rm s}}{\bar{T}^{d-2}}\right)^{\frac{2}{d-2}}
\frac{1}{c_{\rm s}^2}\bar{u}^{\mu} \partial_\mu (\partial_\la\partial^\la \delta\psi),\ee
where $\partial_\mu \psi =h u_\mu \propto T u_\mu$, with $h$ the enthalpy density and then in the plane wave approximation the dispersion relation is
\be\label{frequencyft} \omega= \pm c_{\rm s} k-i\Gamma_{\rm s} \vec{k}^2+\mathcal{O}(k^3).\ee
where we have set the background fluid velocity $\vec{\bar{u}}=0$ in the above equation, and
\be\label{gamma} \Gamma_{\rm s} = \left(\frac{d-2}{d-1}\right)\frac{\eta}{\bar{T} \bar{s}}= \left(\frac{d-2}{d-1}\right)\frac{\eta}{\bar{\epsilon}+\bar{p}}
\ee
is called the attenuation constant which characterize the dissipation of the fluid. The dispersion relation eq.(\ref{frequencyft}) is in accord with the result obtained from doing linearized perturbation in the static conformal fluid \cite{Baier:2007ix}.

%%%%%%%%%%%%%%%%%%%%%%%%%%%%%%%%%%%%%%%%%%%%%%%%%%%%%%%%%%%%%%%%%%%%%%
\section{Holographic derivation of acoustic black holes}\label{sec:holo acoustic}
%%%%%%%%%%%%%%%%%%%%%%%%%%%%%%%%%%%%%%%%%%%%%%%%%%%%%%%%%%%%%%%%%%%%%%
Although the acoustic black hole formed from the supersonic phenomena in hydrodynamics discussed in Sec.\ref{sec:NR acoustic} and Sec.\ref{sec:Rela acoustic} shared similar properties as those of the real black hole, it can only be treated as a black hole analogy. Since their dynamic origins seem to have no relationship with each other. However, as we will show in the rest part of the paper, the acoustic black hole formed in the fluid, can indeed be mapped to a real black hole in an asymptotically AdS spacetime, based on the fluid/gravity duality. Let's consider the bulk $d+1$ dimensional boosted asymptotic AdS
black brane with constant $d$-velocities $u_\mu=\gamma(1,\vec{v})=(u_0,\vec{u})$ and $\eta^{\mu\nu}u_\mu u_\nu=-1$
\be\label{b1} ds^2=H_2(r)
dr^2+H_1(r)\left(-f(r)u_{\mu}u_{\nu}+P_{\mu\nu}\right)dx^{\mu}dx^{\nu},\ee
where the horizon is located at $r=r_{\rm h}$ in which $f(r_{\rm h})=0$ is
satisfied and $P_{\mu\nu}=u_{\mu}u_{\nu}+\eta_{\mu\nu}$ is the
projecting operator. The boosted black brane solution is obtained by making the Lorentz boost transformation $x'^{\mu}=L^{\mu}_{\nu}x^{\nu}$ to the original static black brane
\be\label{bb} ds^2=H_2(r)
dr^2+H_1(r)\left(-f(r)dt^2+dx_i^2\right),\ee
where
\be \label{LTboost}L^0_0=\gamma\equiv u_0,\quad L^0_i=\gamma \beta_i\equiv u_i,
\quad L_{ij}=(\gamma-1)\frac{\beta_i\beta_j}{\vec{\beta}^2}+\delta_{ij}=(\gamma-1)\frac{u^i u_j}{\vec{u}^2 }+\delta_{ij},\quad \vec{\beta}^2=\beta^k\beta_k.\ee
\omits{then
\be\label{bb2} ds^2&=&H_2(r)
dr^2+H_1(r)\left(-f(r)L^0_\mu L^0_\nu dx^{\mu}dx^{\nu}+\delta_{ij} L^i_\mu L^j_\nu dx^{\mu}dx^{\nu}\right)\nno\\
&=&H_2(r)
dr^2+H_1(r)\left(-f(r)u_\mu u_\nu dx^{\mu}dx^{\nu}+\delta_{ij} L^i_0 L^j_0 dt^2+2\delta_{ij} L^i_0 L^j_k dt dx^k+\delta_{ij} L^i_k L^j_l dx^k dx^l\right)\nno\\
&=&H_2(r)
dr^2+H_1(r)\bigg(-f(r)u_\mu u_\nu dx^{\mu}dx^{\nu}+\delta_{ij} u^i u^j dt^2+2\delta_{ij} u^i \left((\gamma-1)\frac{u^j u_k}{\vec{u}^2}+\delta^j_k\right) dt dx^k\nno\\
&&+\delta_{ij}\left((\gamma-1)\frac{u^i u_k}{\vec{u}^2}+\delta^i_k\right)\left((\gamma-1)\frac{u^j u_l}{\vec{u}^2}+\delta^j_l\right)dx^k dx^l
\bigg)\nno\\
&=&H_2(r)
dr^2+H_1(r)\bigg(-f(r)u_\mu u_\nu dx^{\mu}dx^{\nu}+\delta_{ij} u^i u^j dt^2+2u_0 u_k dt dx^k\nno\\
&&+\delta_{ij}\left((\gamma-1)^2\frac{u^i u_k u^j u_l}{\vec{u}^4}+\delta^i_k(\gamma-1)\frac{u^j u_l}{\vec{u}^2}+\delta^j_l(\gamma-1)\frac{u^i u_k}{\vec{u}^2}+\delta^i_k\delta^j_l\right)dx^k dx^l
\bigg)\nno\\
&=&H_2(r)
dr^2+H_1(r)\bigg(-f(r)u_\mu u_\nu dx^{\mu}dx^{\nu}+\delta_{ij} u^i u^j dt^2+2u_0 u_k dt dx^k\nno\\
&&+\left((\gamma-1)^2\frac{u_k u_l}{\vec{u}^2}+2(\gamma-1)\frac{u_k u_l}{\vec{u}^2}+\delta_{kl}\right)dx^k dx^l
\bigg)\nno\\
&=&H_2(r)
dr^2+H_1(r)\bigg(-f(r)u_\mu u_\nu dx^{\mu}dx^{\nu}+\delta_{ij} u^i u^j dt^2+2u_0 u_k dt dx^k+\left((\gamma^2-1)\frac{u_k u_l}{\vec{u}^2}+\delta_{kl}\right)dx^k dx^l
\bigg)\nno\\
&=&H_2(r)
dr^2+H_1(r)\bigg(-f(r)u_\mu u_\nu dx^{\mu}dx^{\nu}+\delta_{ij} u^i u^j dt^2+2u_0 u_k dt dx^k+\left(u_k u_l+\delta_{kl}\right)dx^k dx^l
\bigg)\nno\\
&=&H_2(r)
dr^2+H_1(r)\bigg(-f(r)u_\mu u_\nu dx^{\mu}dx^{\nu}+P_{\mu\nu}dx^\mu dx^\nu
\bigg),\ee
namely, $\delta_{ij} L^i_0 L^j_0=P_{00}$, $\delta_{ij} L^i_0 L^j_k=P_{0k}$ and $\delta_{ij} L^i_k L^j_l=P_{kl}$.}

To remove the coordinate singularity at the horizon, eq.(\ref{b1}) can be written in the Eddington-Finkelstein coordinate as
\be\label{b2}
ds^2=\pm 2\sqrt{H_1(r)H_2(r)f(r)}u_\mu dx^\mu dr+H_1(r)\left(-f(r)u_\mu u_\nu +P_{\mu\nu}dx^{\mu}dx^{\nu}\right)\ee
via the coordinate transformation
\be\label{outgoing-ef} dx^\mu\rightarrow dx'^\mu=dx^\mu \pm u^\mu dr_* = dx^\mu \pm u^\mu \sqrt{\frac{H_2(r)}{H_1(r)f(r)}}dr,\ee
in which $r_*$ is the tortoise coordinate, and $``+"$ indicates the outgoing while $``-"$ corresponds to the ingoing coordinates. Since our purpose is to construct the holographic acoustic black hole on the membrane moving between the bulk black brane horizon or stretched horizon and the asymptotical boundary, so in the following analysis, we can choose the boosted black brane metric either in eq.(\ref{b1}) or in eq.(\ref{b2}) as the bulk background.

From the fluid/gravity duality, the temperature $T$ and entropy $S$ of the fluid are identical to the Hawking temperature $T_{\rm H}$ and entropy $S_{\rm BH}$ of the dual bulk black brane, which are respectively
\be T=T_{\rm H}=\frac{f'\sqrt{H_1}}{4\pi \sqrt{f H_2}}\bigg|_{r_{\rm h}}\quad {\rm and}\quad S=S_{\rm BH}=\frac{\left(H_1(r_{\rm h})\right)^{d-1}}{4G_{d+1}}\int d^{d-1}x.\ee
At the cutoff surface $r=r_{\rm c}$, the reduced fluid velocity contains a redshift
factor as \cite{Emparan:2012be} (note that when the hydrodynamic fluctuations are taken into account, the fluid velocity can become slowly varying functions with respect to the spacetime)
\be \tilde{u}_\mu(r_{\rm c})=\sqrt{f(r_{\rm c})}u_\mu, \ee
and the temperature measured by local observers on the membrane $r=r_{\rm c}$ is
\be T_{\rm c}=\frac{T_{\rm H}}{\sqrt{f(r_{\rm c})}}\geq T_{\rm H}.
\ee
For the asymptotical neutral AdS$_{d+1}$ black brane
we have
\be \label{Dbrane}H_1(r)=\frac{r^2}{L^2},\quad f(r)=1-\frac{r_{\rm h}^d}{r^d} \quad {\rm and}\quad H_2(r)=\frac{1}{f(r)H_1(r)},\ee
then the temperature and the entropy volume density of the dual boundary field theory are
\be
T_{\rm H}=\frac{r_{\rm h} d}{4\pi L^2}\quad {\rm and}\quad
s=\frac{r_{\rm h}^{d-1}}{4 G_{d+1}L^{d-1}}=\frac{1}{4 G_{d+1}}\left(\frac{4\pi T_{\rm H} L}{d}\right)^{d-1},\ee
respectively. In addition, from the first law of thermodynamics
\be d\epsilon=T ds,\ee
the energy density of the dual boundary field theory is
\be \epsilon=\frac{(d-1)r_{\rm h}^d}{16\pi G_{d+1} L^{d+1}},\ee
and the Euler relation
\be \epsilon+p=T s \quad {\rm with }\quad p=\frac{r_{\rm h}^d}{16\pi G_{d+1} L^{d+1}} \ee
is satisfied at the asymptotical boundary of the AdS$_{d+1}$ spacetime.

%%%%%%%%%%%%%%%%%%%%%%%%%%%%%%%%%%%%%%%%%%%%%%%%%%%%%%%%%%%%%%%%%%%%%%
\subsection{Stress tensor of the fluid at cutoff surfaces}
%%%%%%%%%%%%%%%%%%%%%%%%%%%%%%%%%%%%%%%%%%%%%%%%%%%%%%%%%%%%%%%%%%%%%%
The renormalized holographic stress tensor $T_{{\rm c}\mu\nu}$ on the cutoff surface $\Si_{\rm c}(r=r_{\rm c})$ can be obtained from the Brown-York formalism \cite{Henningson:1998gx,Balasubramanian:1999re,de Haro:2000xn}, for the spacetime background in eq.(\ref{b1}), it is
\be T_{{\rm c}\mu\nu}=-\frac{H_1^{\frac{d}{2}}}{8\pi
G_{d+1}}\left(\hat{K}_{{\rm c}\mu\nu}-\hat{\gamma}_{\mu\nu}\hat{K}_{\rm c}+(d-1)\frac{\hat{\gamma}_{\mu\nu}}{L}+\cdots\right),\ee
where the induced extrinsic curvature
\be
\hat{K}_{{\rm c}\mu\nu}=\frac{1}{2
H_1(r_{\rm c})}\frac{1}{\sqrt{H_2(r_{\rm c})}}\partial_r \gamma_{\mu\nu}(r_{\rm c}),
\ee
and $\gamma_{\mu\nu}(r_{\rm c})=H_1(r_{\rm c})\hat{\gamma}_{\mu\nu}$ is the induced
metric on the timelike cutoff surface and $T_{\rm c}^{\mu\nu}=\gamma^{\mu\alpha}\gamma^{\nu\beta}T_{{\rm c}\alpha\beta}$. While ``$\cdots$'' represent higher derivative terms constructed from the induced metric in order
to cancel the UV divergences and theses higher curvature terms vanish for the spatially flat case considered here.

Let's consider the bulk neutral black AdS$_{d+1}$ brane, using eq.(\ref{Dbrane}), then the
holographic stress tensor at the cutoff surface can be expressed as that for the ideal fluid \cite{Emparan:2013ila}
\be\label{sTc} T_{{\rm c}\mu\nu}=\epsilon_{\rm c} \tilde{u}_\mu \tilde{u}_\nu + p_{\rm c}
P_{\mu\nu},\ee
where energy density and pressure at the cutoff surface are respectively
\be \epsilon_{\rm c}=\frac{(d-1)r_{\rm c}^d}{8\pi G_{d+1} L^{d+1}}\left(1-\sqrt{f(r_{\rm c})}\right)\quad {\rm and}\quad
p_{\rm c}=-\epsilon_{\rm c}+\frac{r_{\rm h}^d d}{16\pi G_{d+1} L^{d+1}
\sqrt{f(r_{\rm c})}},\ee
which means that the Euler relation
\be\label{Euler} T_{\rm c} s_{\rm c}=\epsilon_{\rm c}+p_{\rm c}\ee
and the first law of thermodynamics
\be\label{1stlawcut}\delta \epsilon_{\rm c} = T_{\rm c} \delta s_{\rm c},\ee
are still held at the cutoff surface $r=r_{\rm c}$ when the fluid is isentropic,
where the entropy density $s_{\rm c}$ of fluid at the cutoff surface is the same as that of the fluid at the AdS boundary, i.e. $s_{\rm c}=s$ and the variation $\delta$ is acting on the horizon radius $r_{\rm h}$. Then the speed of sound of the fluid at the cutoff surface can be computed directly via
\be\label{sound} \hat{c}_{\rm s}^2 &=& \frac{\delta p_{\rm c}}{ \delta \epsilon_{\rm c}}\bigg|_{r=r_{\rm c}}=\frac{1}{d-1} + \frac{d}{2(d-1)}\frac{r_{\rm h}^d}{r_{\rm c}^d f(r_{\rm c})}\nno\\
&=& c_\infty^2 + \frac{d}{2(d-1)}\frac{1-f(r_{\rm c})}{f(r_{\rm c})},\ee
where $c_\infty=\sqrt{1/(d-1)}$ is the value of the speed of sound of the $d$-dimensional conformal fluid. It can be seen that the speed of sound is a monotonically decreasing function of $r_{\rm c}$ which runs along the radial direction from the value $\hat{c}_{\rm s}=\infty$ at the black hole horizon $r_{\rm c}=r_{\rm h}$ to $\hat{c}_{\rm s}=\sqrt{1/(d-1)}$ at $r_{\rm c}\rightarrow \infty$, which indicates that the fluid is incompressible when the cutoff surface or the membrane is chosen at the horizon while it becomes compressible when $r_{\rm c}>r_{\rm h}$.
%(Since the speed of sound is running with respect to the radial coordinate, for the fluid located at the cutoff surface with constant velocities, there may exist a region $r\geq r_{\rm eq}$ such that the $c_c\leq u$.) Note that when the bulk dual black brane contains charges, $c_c$ has a quadratic pole in the near horizon near extreme region, in that case, it is possible to obtain the corresponding near extreme acoustic black hole near the bulk black brane horizon.

%%%%%%%%%%%%%%%%%%%%%%%%%%%%%%%%%%%%%%%%%%%%%%%%%%%%%%%%%%%%%%%%%%%%%%
\subsection{Acoustic black hole from fluid at the cutoff surface}\label{subsec:holoacoustic}
%%%%%%%%%%%%%%%%%%%%%%%%%%%%%%%%%%%%%%%%%%%%%%%%%%%%%%%%%%%%%%%%%%%%%%
From the fluid/gravity duality, the conservation equation of the fluid at the cutoff surface is equivalent to the constraint equation of $r\mu$ components of the bulk Einstein equation. Then the acoustic metric of the fluid at the cutoff surface can be constructed from perturbing the longitudinal mode of the conservation equation $\tilde{u}^\nu \nabla_{{\rm c}\mu} T_{{\rm c}\nu}^{\mu}=0$,
\be \label{longmode}
\nabla_{{\rm c}\mu}\left(\epsilon_{\rm c} \tilde{u}^\mu\right)+p_{\rm c}\nabla_{{\rm c}\mu}\left(\tilde{u}^\mu\right)=0,\ee
and the transverse mode $P^\nu_{\lambda} \nabla_{{\rm c}\mu} T_{{\rm c}\nu}^{\mu}=0$, i.e.
\be\label{transmode} P^{\mu}_\la\nabla_{{\rm c}\mu} p_{\rm c}+\left(\epsilon_{\rm c}+p_{\rm c}\right)\tilde{u}^\mu\nabla_{{\rm c}\mu}\tilde{u}_\la=0,\ee
where $\nabla_{{\rm c}\mu}$ is the covariant derivative compatible with the induced metric $\gamma_{\mu\nu}$ at the cutoff surface and we have used the normalized condition $\gamma^{\mu\nu}\tilde{u}_\mu\tilde{u}_\nu=-1$. In addition, we have required that the stress tensor $T_{\rm c}^{\mu\nu}$ at the cutoff surface still to have the form of the ideal fluid, although $\tilde{u}_\nu$, $\epsilon_{\rm c}$ and $p_{\rm c}$ became slowly varying functions of $x^\alpha$. Generally speaking, from the bulk gravity side, such kind of hydrodynamic fluctuations will cause $T_{\rm c}^{\mu\nu}$ to be modified by the dissipative terms. However, as we have discussed in Sec.\ref{sec:Rela acoustic}, the presence of viscosities will only modify the EoM for the phonon by the third derivative terms and thus break the Lorentz symmetry of the fluid, while the acoustic geometry remains the same as that of the ideal fluid. Thus at this step, we can still use the EoMs of the ideal fluid, i.e. eqs.(\ref{longmode})(\ref{transmode}) to determine the acoustic geometry.

Following the similar steps as those in the relativistic irrotational fluid \cite{Visser:2010xv}, namely, defining
\be\label{vpotential}
u=u_\mu dx^{\mu} \equiv \frac{\nabla_{{\rm c}\mu} \psi}{\sqrt{-\gamma^{\alpha\beta}\nabla_{{\rm c}\alpha}\psi\nabla_{{\rm c}\beta}\psi}}dx^{\mu}
 \ee
and the vorticity free condition can be expressed as
\be \tilde{u}\wedge d\tilde{u}=0,\ee
which reduces to the non-relativistic one when the velocity does not depend on time apparently.

Perturbing the fluid up to the linearized order as
\be\label{normmode} \psi=\bar{\psi}+\delta\psi,\quad \epsilon_{\rm c}=\bar{\epsilon}_{\rm c}+\delta\epsilon_{\rm c}\quad {\rm and} \quad p_{\rm c}=\bar{p}_{\rm c}+\delta p_{\rm c}.\ee
Then the EoM for the phonon, i.e. the perturbation of the velocity potential, is
\be\label{eomcut} \partial_\mu\left(\frac{\bar{n}_{\rm c}^2}{\bar{\epsilon}_{\rm c}+\bar{p}_{\rm c}}
\left(-\frac{1}{\hat{c}_{\rm s}^2}\bar{\tilde{u}}^\mu\bar{\tilde{u}}^\nu+P^{\mu\nu}\right)\partial_\nu\delta\psi\right)=0.
\ee
Consequently, the relativistic acoustic metric at the cutoff surface can be obtained
\be\label{accut} ds^2_{\rm ac}=\left(\frac{\bar{n}_{\rm c}^2 }{\hat{c}_{\rm s}\left(\bar{\epsilon}_{\rm c}+\bar{p}_{\rm c}\right)}\right)^{\frac{2}{d-2}}
\left(-\hat{c}_{\rm s}^2 \bar{\tilde{u}}_{\mu}\bar{\tilde{u}}_{\nu} + P_{\mu\nu}\right)dx^\mu dx^\nu,\ee
where $\bar{n}_{\rm c}$ is the zeroth order particle number density of the fluid which is proportional to the fluid entropy density and
\be n_{\rm c}=n_0\exp{\int\frac{d\epsilon_{\rm c}}{\epsilon_{\rm c}+p_{\rm c}}},\ee
where $n_0\equiv n_{\rm c}(p_{\rm c}=0)$. When approaching the asymptotical boundary $r_{\rm c}\rightarrow \infty$, we have
\be \epsilon_\infty&=&\frac{(d-1)r_{\rm c}^d}{8\pi G_{d+1} L^{d+1}}\left(1-\sqrt{f(r_{\rm c})}\right)\bigg|_{r_{\rm c}\rightarrow\infty}\rightarrow \frac{(d-1)r_{\rm h}^d}{16\pi G_{d+1} L^{d+1}},\nno\\
p_\infty&=&-\epsilon_{\rm c}+\frac{r_{\rm h}^d d}{16\pi G_{d+1} L^{d+1}
\sqrt{f(r_{\rm c})}}\bigg|_{r_{\rm c}\rightarrow\infty}\rightarrow \frac{r_{\rm h}^d }{16\pi G_{d+1} L^{d+1}},\ee
then eq.(\ref{accut}) reduces to the form in the flat spacetime case
\be\label{acinfi} ds^2_{\rm ac}&=&\epsilon_\infty^{\frac{2}{d}}\left(\frac{n_0^2 }{c_\infty^3 d}\right)^{\frac{2}{d-2}}
\left(-c_\infty^2 \bar{u}_{\mu}\bar{ u}_{\nu} + P_{\mu\nu}\right)dx^\mu dx^\nu\nno\\
&=&T^2\left(\frac{n_0^2 \sigma^{\frac{d-2}{d}} }{c_\infty^3 d}\right)^{\frac{2}{d-2}}
\left(-c_\infty^2 \bar{u}_{\mu}\bar{ u}_{\nu} + P_{\mu\nu}\right)dx^\mu dx^\nu,\ee
as in eq.(\ref{acour}), where $\sigma=\frac{1}{4G_{d+1}}\left(\frac{4\pi L}{d}\right)^{d-1}$. More explicitly, the acoustic geometry eq.(\ref{accut}) can be written in a form as
\be\label{accut2}ds_{\rm ac}^2=\left(\frac{\bar{n}_{\rm c}^2 }{\hat{c}_{\rm s}\left(\bar{\epsilon}_{\rm c}+\bar{p}_{\rm c}\right)}\right)^{\frac{2}{d-2}}\left(-\left(1-\gamma^2(1-\hat{c}_{\rm s}^2 f(r_{\rm c}))\right)d\tau^2+\left(\delta_{ij}+\frac{\left(\frac{1}{f(r_{\rm c})}-\hat{c}_{\rm s}^2\right)\bar{\tilde{u}}_{i}\bar{\tilde{u}}_{j}}{1-\gamma^2(1-\hat{c}_{\rm s}^2 f(r_{\rm c}))}\right)dx^i dx^j\right),\nno\\
\ee
in which
\be
dt=d\tau-\frac{\left(\hat{c}_{\rm s}^2-\frac{1}{f(r_{\rm c})}\right)\bar{\tilde{u}}_{0}\bar{\tilde{u}}_{i}}{1+\gamma^2\left(\hat{c}_{\rm s}^2 f(r_{\rm c})-1\right)}dx^i.
\ee
Without loss of generality, one can choose the coordinate to let the fluid flowing along the $x^{d-1}\equiv z$ direction, namely, $\bar{u}_\mu=(\gamma,0,\ldots,0,\bar{u}_z)=\gamma(1,0,\ldots,0,\bar{v}_z)$. Thus eq.(\ref{accut2}) reduces to
\be\label{accut3}ds_{\rm ac}^2=\left(\frac{\bar{n}_{\rm c}^2 }{\hat{c}_{\rm s}\left(\bar{\epsilon}_{\rm c}+\bar{p}_{\rm c}\right)}\right)^{\frac{2}{d-2}}\left(
-\left(f(r_{\rm c})\hat{c}_{\rm s}^2-\bar{\tilde{u}}_z^2\al_{\rm c}^2 \right)d\tau^2 +\frac{f(r_{\rm c})\hat{c}_{\rm s}^2}{f(r_{\rm c})\hat{c}_{\rm s}^2-\bar{\tilde{u}}_z^2\al_{\rm c}^2}dz^2 +dx^a dx_a\right),
\ee
where we have defined
\be
\al_{\rm c}^2\equiv\frac{1}{f(r_{\rm c})}-\hat{c}_{\rm s}^2=\frac{(d-2)}{2(d-1)}\left(1+\frac{1}{f(r_{\rm c})}\right),
\ee
Then it is straightforward to check that supersonic phenomena appears or the acoustic black hole forms when
\be\label{supersonic} f(r_{\rm c})\hat{c}_{\rm s}^2-\bar{\tilde{u}}_z^2\al_{\rm c}^2\leq 0,\ee
which gives
\be \bar{\tilde{u}}_z\geq \frac{\sqrt{f(r_{\rm c})}\hat{c}_{\rm s}}{\al_{\rm c}}\quad {\rm or}\quad  \bar{\tilde{u}}^z=\frac{\bar{u}^z}{\sqrt{f(r_{\rm c})}}\geq \frac{\bar{v}_{z}}{\sqrt{f(r_{\rm c})}}\geq \hat{c}_{\rm s}.\ee
For example, when the cutoff surface is taken to the AdS boundary, from eq.(\ref{supersonic}) we have
\be
\bar{u}_{z} \geq \frac{1}{\sqrt{d-2}} \quad {\rm or}\quad \bar{v}_{z}\geq \frac{1}{\sqrt{d-1}} = c_\infty.\ee

Furthermore, we can write the acoustic metric eq.(\ref{accut3}) in a more explicit form by requiring $\bar{\tilde{u}}_{z}=\bar{\tilde{u}}_{z}(z)$ to be a monotonically increasing function of $z$ (with direction along $-z$) and it reaches the critical value at the acoustic horizon $\bar{\tilde{u}}_{z}(z_{\rm sh})=\frac{\sqrt{f(r_{\rm c})}\hat{c}_{\rm s}}{\al_{\rm c}}\equiv \frac{\tilde{\hat{c}}_{\rm s}}{\al_{\rm c}}$. Then expand $\bar{\tilde{u}}_{z}(z)$ around the spacetime point $(t=0,\ldots,0,z_{\rm sh})$ up to the linear order in a covariant form, namely, $\bar{\tilde{u}}_{z}(z)=\bar{\tilde{u}}_{z}(z_{\rm sh})+(z-z_{\rm sh})\partial_{z} \bar{\tilde{u}}_{z}|_{z_{\rm sh}}$, eq.(\ref{accut3}) becomes
\be\label{accut4}ds_{\rm ac}^2 = \left(\frac{\bar{n}_{\rm c}^2 }{\hat{c}_{\rm s}\left(\bar{\epsilon}_{\rm c}+\bar{p}_{\rm c}\right)}\right)^{\frac{2}{d-2}}\left(
-2\al_{\rm c} \tilde{\hat{c}}_{\rm s}|\partial_{z} \bar{\tilde{u}}_{z}|\left(z-z_{\rm sh}\right) d\tau^2 +\frac{\tilde{\hat{c}}_{\rm s}}{2\al_{\rm c} |\partial_{z} \bar{\tilde{u}}_{z}|\left(z-z_{\rm sh}\right)}dz^2+ dx^a dx_a\right)
\ee
and the temperature of the acoustic horizon is
\be \label{Tsre}
T_{\rm sh}=\frac{\al_{\rm c}}{2\pi}|\partial_{z} \bar{\tilde{u}}_{z}|_{z_{\rm sh}},\ee
which is related to the acceleration of the background fluid at the acoustic horizon $z=z_{\rm sh}$ as
\be\label{acac} \bar{\tilde{a}}_{z}|_{z_{\rm sh}}=\bar{\tilde{u}}^\al\nabla_{{\rm c}\al}\bar{\tilde{u}}_{z}|_{z_{\rm sh}}=-\frac{2\pi\hat{c}_{\rm s}T_{\rm sh}}{\al_{\rm c}^2\sqrt{f(r_{\rm c})}}+\partial_z\ln\sqrt{f(r_{\rm c})}|_{z_{\rm sh}}.
\ee
For convenience, we will omit the ``bar'' index for the background hydrodynamic variables in the following subsections.

%%%%%%%%%%%%%%%%%%%%%%%%%%%%%55%%%%%%%%%%%%%%%%%%%%%%%%%%%%%%%%%%%%%%%
\subsection{Perturbations from the bulk side}\label{subsec:hydroperturb}
%%%%%%%%%%%%%%%%%%%%%%%%%%%%%%%%%%%%%%%%%%%%%%%%%%%%%%%%%%%%%%%%%%%%%%
Let us study the holographic dual of the acoustic black hole in eq.(\ref{accut4}) from the perturbations of the bulk black brane in eq.(\ref{b1}) caused by small hydrodynamic fluctuations at the cutoff surface, and we will focus on the linearized perturbation. Recall that hydrodynamics describes states close to the thermal equilibrium, which allows the local velocity fields as well as the local temperature of the fluid to be slowly varying functions of the spacetime coordinates $x^\al$, namely~\cite{Bhattacharyya:2008jc,Kovtun:2012rj},
\be u^\mu\rightarrow u^\mu(x^\al)\quad {\rm and}\quad T\rightarrow T(x^\al) \quad ({\rm or} \quad r_{\rm h}\rightarrow r_{\rm h}(x^\al)).\ee
These hydrodynamic fluctuations can be viewed as the variation of source terms from the fluid, which will in turn cause backreaction to the background geometry, which indicates that, the acoustic black hole formed in the fluid at the cutoff surface is corresponds to the bulk asymptotic AdS black brane with first order corrections. In order to make the perturbed geometry to satisfy the original bulk Einstein equation, additional metric corrections $g^{(1)}_{AB}$ should be added to the perturbed metric. To solve the fluctuations, it's more convenient to use the metric in the ingoing Eddington-Finkelstein coordinate and make a scaling transformation for the time coordinate $t\rightarrow t/\sqrt{f(r_{\rm c})}$, then eq.(\ref{b2}) becomes to
\be \label{b4}
ds^2&=&- 2 \frac{u_\mu(x^\alpha)}{\sqrt{f\left(r_{\rm c},r_{\rm h}(x^\alpha)\right)}} d x^\mu d r +\frac{r^2}{L^2}\left(\left(1-\frac{f\left(r,r_{\rm h}(x^\alpha)\right)}{f\left(r_{\rm c},r_{\rm h}(x^\alpha)\right)}\right) u_\mu(x^\alpha) u_\nu(x^\alpha)+\eta_{\mu\nu}\right) d x^\mu d x^\nu\nno\\
&&+ \left(g^{(1)}_{\mu\nu}(r,x^\alpha)dx^\mu dx^\nu + 2 g^{(1)}_{\mu r}(r,x^\alpha)dx^\mu dr + g^{(1)}_{rr}(r,x^\alpha)dr^2\right),\ee
where we have chosen the background to be the AdS$_{d+1}$ black brane. To compare eq.(\ref{b4}) with its holographic counterpart in eq.(\ref{accut4}), we will rotate the the coordinates to let the fluid moving in the $-z$ direction and further require that the local velocity fields and the local temperature of the fluid only to be the function of the coordinate $z$. Then the vorticity free condition $u\wedge du=0$ is automatically satisfied. In addition, since we are interested in the phenomena in the near acoustic horizon region, we will expand $u_{z}(z)$ and $r_{\rm h}(z)$ at the location of the acoustic horizon $(t=0,\cdots,0,z=z_{\rm sh})$ without loss of generality, this is similar with the usual treatments in the fluid/gravity duality in which the hydrodynamic variables are expanded at $x^\al=0$. Therefore, the non-vanishing metric components (denoted by $g^{(o)}_{AB}$) in the first line of eq.(\ref{b4}) are
\be\label{b5}
g^{(o)}_{tt}&=& \frac{r^2}{L^2}\left[u_t^2(z) \left(1-\frac{f\left(r, r_{\rm h}(z)\right)}{f\left(r_{\rm c}, r_{\rm h}(z)\right)}\right) -1\right],\nno\\
g^{(o)}_{tz}&=& g^{(o)}_{zt}=\frac{r^2}{L^2}u_t(z) u_z(z) \left(1-\frac{f\left(r, r_{\rm h}(z)\right)}{f\left(r_{\rm c}, r_{\rm h}(z)\right)}\right),\nno\\
g^{(o)}_{ab}&=& \frac{r^2}{L^2}\delta_{ab}\quad ({\rm with}\quad a,b\neq z),\nno\\
g^{(o)}_{zz}&=& \frac{r^2}{L^2}\left[u_z^2(z) \left(1-\frac{f\left(r, r_{\rm h}(z)\right)}{f\left(r_{\rm c}, r_{\rm h}(z)\right)}\right) +1\right],\nno\\
g^{(o)}_{rt}&=& g^{(o)}_{tr}= -\frac{u_t(z)}{\sqrt{f\left(r_{\rm c}, r_{\rm h}(z)\right)}},\nno\\
g^{(o)}_{rz}&=& g^{(o)}_{zr}= -\frac{u_z(z)}{\sqrt{f\left(r_{\rm c}, r_{\rm h}(z)\right)}},
\ee
where
\be u_{z}(z) &=& \frac{\hat{c}_{\rm s}}{\al_{\rm c}}+(z-z_{sh})\left(-\frac{2\pi T_{\rm sh}}{\al_{\rm c}\sqrt{f(r_{\rm c})}}+\frac{d\hat{c}_{\rm s}r_{\rm h}^{d-1}}{2\al_{\rm c}f(r_{\rm c})r^d}\partial_z r_{\rm h}|_{z_{\rm sh}}\right),\nno\\
u_{t}(z) &=& \frac{1}{\al_{\rm c}f(r_{\rm c})}+(z-z_{sh})\left(-\frac{2\pi\hat{c}_{\rm s} T_{\rm sh}}{\al_{\rm c}}+\frac{d\hat{c}_{\rm s}^2r_{\rm h}^{d-1}}{2\al_{\rm c}\sqrt{f(r_{\rm c})}r^d}\partial_z r_{\rm h}|_{z_{\rm sh}}\right),\nno\\
r_{\rm h}(z) &=& r_{\rm h}(z_{\rm sh})+(z-z_{\rm sh})\partial_z r_{\rm h}|_{z_{\rm sh}}\equiv r_{\rm h}+(z-z_{\rm sh})\partial_z r_{\rm h}|_{z_{\rm sh}},\nno\\
f\left(r,r_{\rm h}(z)\right)&=& f\left(r,r_{\rm h}\right)-\frac{d r_{\rm h}^{d-1}(z-z_{\rm sh})\partial_z r_{\rm h}|_{z_{\rm sh}}}{r^d}\equiv f(r)-\frac{d r_{\rm h}^{d-1}(z-z_{\rm sh})\partial_z r_{\rm h}|_{z_{\rm sh}}}{r^d},\nno\\
f\left(r_{\rm c},r_{\rm h}(z)\right)&=& f\left(r_{\rm c},r_{\rm h}\right)-\frac{d r_{\rm h}^{d-1}(z-z_{\rm sh})\partial_z r_{\rm h}|_{z_{\rm sh}}}{r_{\rm c}^d}\equiv f(r_{\rm c})-\frac{d r_{\rm h}^{d-1}(z-z_{\rm sh})\partial_z r_{\rm h}|_{z_{\rm sh}}}{r_{\rm c}^d},
\ee
which contain both the zeroth and the first order derivative terms of hydrodynamic fluctuations.

On the other hand, the first order metric corrections in the second line of eq.(\ref{b4}) can be decomposed into the $SO(d-1)$ algebraically symmetric forms, after further imposing the radial gauge $g^{(1)}_{rA}=0$, they can be expressed into four independent parts as
\be\label{mecor1st}
g^{(1)}_{rr}(r)&=& g^{(1)}_{\mu r}(r,x^\alpha)=0,\nno\\
g^{(1)}_{\mu \nu}(r,x^\alpha)&=& \theta u_\mu u_\nu \mathfrak{s}_1(r)+\frac{\theta}{d-1}P_{\mu\nu}\mathfrak{s}_2(r)
+2a_{(\mu}u_{\nu)}\mathfrak{v}(r)+\sigma_{\mu\nu}\mathfrak{t}(r),
\ee
in which
\be\label{uasigcfm} \theta=\partial_\mu u^\mu,\quad a_\mu=u^\nu\partial_\nu u_\mu \quad {\rm and}\quad
\sigma_{\mu\nu}=P_\mu^\alpha P_\nu^\beta \partial_{(\alpha}u_{\beta)}-\frac{\theta}{d-1}P_{\mu\nu}\ee
are respectively the expansion, acceleration and shear tensor associated with the velocity field $u_\mu$ of the dual fluid, and the index are lowered and raised by $\eta_{\mu\nu}$ and $\eta^{\mu\nu}$, since the reduced metric on the cutoff surface is flat in the rescaled coordinates. The functions $\mathfrak{s}_i(r)$ ($i=1,2$), $\mathfrak{v}(r)$ and $\mathfrak{t}(r)$ belong to the scalar, vector and tensor channels, respectively and will decouple with each other in the linearized Einstein's equation. To determine the first order perturbed bulk geometry, we need to substitute eqs.(\ref{b5})(\ref{mecor1st}) into eq.(\ref{b4}) and solve the linearized bulk Einstein's equation
\be\label{Einstein1st}\delta R_{AB}=-\frac{d}{L^2}\delta g_{AB}.\ee
A natural boundary condition for solving the metric corrections are the Dirichlet boundary condition at the cutoff surface, i.e. $\mathfrak{s}_1(r_{\rm c})=\mathfrak{s}_2(r_{\rm c})=\mathfrak{v}(r_{\rm c})=\mathfrak{t}(r_{\rm c})=0$. What is more, additional gauge should be adopted to solve eq.(\ref{Einstein1st}), and we use the Landau frame, in which the first order correction of the fluid stress tensor is transverse at the timelike cutoff surfaces, i.e.
\be \tilde{u}^{\mu} T^{(1)}_{{\rm c}\mu\nu}=0.
\ee
The general solution has been solved out in \cite{Pinzani-Fokeeva:2014cka} as
\begin{align}
 \mathfrak{s}_1(r) &=\frac{r}{(d-1)f(r_{\rm c})^{3/2}}\left[- d  \left(1-\frac{r_{\rm c}^{d-1}}{r^{d-1}}\right) +(d-2) f(r)
 +\left(2-d\frac{r_{\rm c}^{d-1}}{r^{d-1}}\right)f(r_{\rm c})\right],\nno\\
\mathfrak{s}_2(r)&=\frac{2r}{\sqrt{f(r_{\rm c})}} \left(1-\frac{r}{r_{\rm c}}\right)  \nno\\
\mathfrak{v}(r) &= -\frac{r}{(d-1)\hat{c}_{\rm s}^2 r_{\rm c}f(r_{\rm c})^{3/2}}\left[ r \left(\frac{r_{\rm c}^d}{r^d}-1\right)+r_{\rm c} f(r_{\rm c})\left(1-\frac{r_{\rm c}^{d-1}}{r^{d-1}}\right)\right]\nno\\
\mathfrak{t}(r) &=\frac{2}{r_{\rm h}}\sqrt{f(r_{\rm c})} r^2\left[h(r)-h(r_{\rm c})+\frac{1}{d}\ln\frac{f(r)}{f(r_{\rm c})}  \right]\quad {\rm and}\quad h(r)=\frac{r_{\rm h}}{r}{_{2}F_1}\left(1,\frac{1}{d}, 1+\frac{1}{d}, \frac{r_{\rm h}^d}{r^d}\right),
 \end{align}
and the nonvanishing components of $g^{(1)}_{\mu \nu}$ will be listed in Appendix \ref{sec:1st correc}. Besides, the first order corrections of the stress tensor at finite cutoff surface $r=r_{\rm c}$ is given by
\begin{align}
T_{c}^{\mu\nu(1)}=-2\eta_{\rm c} \sigma^{\mu\nu} +\zeta_{\rm c} \theta P^{\mu\nu},\quad \eta_{\rm c}=\frac{r_{\rm h}^{d-1}}{16\pi G_{d+1}L^{d-1}},\quad \zeta_{\rm c}=0,
\end{align}
where $\eta_{\rm c}$ and $\zeta_{\rm c}$ are the shear and bulk viscosities at the cutoff surface, respectively. Note that the shear viscosity and the entropy density of the dual fluid at the rescaled cutoff surface with $x^i\rightarrow x^i L/r_{\rm c}$ are $\eta(r_{\rm c})=\frac{r_{\rm h}^{d-1}}{16\pi G_{d+1}r_{\rm c}^{d-1}}$ and $s(r_{\rm c})=\frac{r_{\rm h}^{d-1}}{4G_{d+1}r_{\rm c}^{d-1}}$, respectively, which both indicate that the shear viscosity over entropy density of the fluid
\be\label{eta/s} \frac{\eta(r_{\rm c})}{s(r_{\rm c})}=\frac{\eta_{\rm c}}{s_{\rm c}}=\frac{1}{4\pi}\ee
will not vary as the cutoff surface $r=r_{\rm c}$ moving from the black brane horizon to the asymptotical boundary of the AdS spacetime.

%%%%%%%%%%%%%%%%%%%%%%%%%%%%%%%%%%%%%%%%%%%%%%%%%%%%%%%%%%%%%%%%%%%%%%
\subsection{Temperature of acoustic black hole vs. temperature of real black hole}
%%%%%%%%%%%%%%%%%%%%%%%%%%%%%%%%%%%%%%%%%%%%%%%%%%%%%%%%%%%%%%%%%%%%%%
Recall that one of the original motivations for studying acoustic black hole was to mimic the Hawking radiation, e.g., detecting the Hawking-like temperature of the acoustic horizon. Nevertheless, the detection of the former cannot indicate the observation of the Hawking temperature of real black holes, in the absence of further evidences such as the dynamical origins. Now using the holographic construction, the temperature of the acoustic black hole can indeed be connected to that of it's dual AdS black brane--a real black hole. Explicitly, the conservation equation $\nabla_{{\rm c}\mu} T_{{\rm c}\nu}^{\mu}=0$, together with the Euler relation eq.(\ref{Euler}) and the first law of thermodynamics eq.(\ref{1stlawcut}), of the fluid at cutoff surface $r=r_{\rm c}$ can be combined into a single equation as
\be\label{Tas}
\nabla_{{\rm c}\mu}\ln s_{\rm c}=\left(\tilde{\theta}\tilde{u}_\mu-\hat{c}_{\rm s}^{-2}\tilde{a}_\mu\right) +\frac{1}{2\pi T_{\rm c}}\left(\tilde{u}_\mu \tilde{u}^\al+\hat{c}_{\rm s}^{-2}P^\al_\mu \right)\nabla_{{\rm c}\beta}\tilde{\sigma}^\beta_\al,
\ee
which can be further written as
\be\label{TaTb}
\nabla_{{\rm c}\mu}\ln r_{\rm h}=\nabla_{{\rm c}\mu}\ln T_{\rm H}=\frac{1}{d-1}\left(\tilde{\theta}\tilde{u}_\mu-\hat{c}_{\rm s}^{-2}\tilde{a}_\mu\right)+\frac{1}{2\pi(d-1) T_{\rm c}}\left(\tilde{u}_\mu \tilde{u}^\al+\hat{c}_{\rm s}^{-2}P^\al_\mu \right)\nabla_{{\rm c}\beta}\tilde{\sigma}^\beta_\al,
\ee
where the expansion is $\tilde{\theta}(x^\al)=\nabla_{{\rm c}\mu}\tilde{u}^\mu(x^\al)$ and acceleration of the fluid velocity $\tilde{u}_\mu(x^\al)$ is $\tilde{a}_\mu(x^\al)=\tilde{u}^\nu(x^\al)\nabla_{{\rm c}\nu}\tilde{u}_\mu(x^\al)$. Since the second part in eq.(\ref{Tas}) or eq.(\ref{TaTb}) are the second order derivative terms coming from the first order dissipations corrections to the fluid, so when the fluid velocity is slowly varying (which requires that the fluid at the cutoff surface is close to thermal equilibrium), namely, its acceleration is small, these subleading terms can be ignored. Besides, using $\nabla_{{\rm c}\mu}\tilde{u}^\nu(x^\al)=\partial_\mu\tilde{u}^\nu(x^\al)+\Gamma_{{\rm c}\mu\la}^\nu \tilde{u}^\la(x^\al)$ (where $\Gamma_{{\rm c}\mu\la}^\nu$ is the Christoffel connection associated with the induced metric $\gamma_{\mu\nu}$ and it will not be modified up to the first order metric corrections due to the Dirichlet boundary condition at $r=r_{\rm c}$), it is straightforward to check that they are respectively related to their counterparts in eq.(\ref{uasigcfm}) as
\be
\tilde{\theta}(x^\al)=\frac{\theta(x^\al)}{\sqrt{f\left(r_{\rm c},r_{\rm h}(x^\al)\right)}},\quad \tilde{a}_\mu(x^\al)=a_\mu(x^\al)+P^\nu_\mu\partial_\nu\ln\sqrt{f\left(r_{\rm c},r_{\rm h}(x^\al)\right)},\quad \tilde{\sigma}^\beta_\al=\frac{\sigma^\beta_\al}{\sqrt{f\left(r_{\rm c},r_{\rm h}(x^\al)\right)}}.
\ee
For the holographic acoustic black hole metric eq.(\ref{accut4}), we obtain the relationship between the temperature of the acoustic black hole $T_{\rm sh}$ and the dual bulk black brane Hawking temperature $T_{\rm H}$ at the acoustic horizon $z=z_{\rm sh}$ as
\be\label{TaTb2}
\partial_z\ln T_{\rm H}|_{z_{\rm sh}}=\frac{4\pi\sqrt{f(r_{\rm c})}(1-\hat{c}_{\rm s}^2)}{2(d-1)f(r_{\rm c})\al_{\rm c}^2-d(1-f(r_{\rm c}))(\hat{c}_{\rm s}^2+\frac{\al_{\rm c}^2}{\hat{c}_{\rm s}^2})}\frac{T_{\rm sh}}{\hat{c}_{\rm s}},
\ee
where the subleading second order derivative terms have been ignored. Eq.(\ref{TaTb2}) has a simple expression when the cutoff surface going to the AdS boundary $r_{\rm c}\rightarrow\infty$, then
\be\label{TaTbbdy}
\partial_z\ln T_{\rm H}|_{z_{\rm sh}}=2\pi T_{\rm sh}c_{\rm s}.
\ee
%

%%%%%%%%%%%%%%%%%%%%%%%%%%%%%%%%%%%%%%%%%%%%%%%%%%%%%%%%%%%%%%%%%%%%%%
\subsection{The sound mode/scalar quasinormal mode duality}
%%%%%%%%%%%%%%%%%%%%%%%%%%%%%%%%%%%%%%%%%%%%%%%%%%%%%%%%%%%%%%%%%%%%%%
Recall that when the spacetime background is the static AdS black brane (or in the locally static frame), the components of metric variation $\delta g_{\mu\nu}(r,x^\alpha)$ (which is the normal mode perturbation with respect to the original unperturbed background geometry eq.(\ref{bb})), if letting the mode propagating along the $z$-coordinate, are $\delta g_{tt}$, $\delta g_{t z}$, $\delta g_{z z}$, $\delta g_{rr}$, $\delta g_{t r}$, $\delta g_{r z}$ and $\delta g^a_{a}$, which will compose the longitudinal channel of the quasinormal mode (which is the $SO(d-2)$ scalar mode) of the bulk gravitational perturbation, this requires the $SO(d-2)$ gauge invariant decomposition of the bulk linearized gravitational equations. Then the dual operators on the AdS boundary corresponding to the rest bulk gravitational perturbations are $T_{\rm c}^{tt}$,  $T_{\rm c}^{t z}$,  $T_{\rm c}^{zz}$ and $T_{{\rm c}a}^{a}$, from the field/operator correspondence in the gauge/gravity duality. In the long wavelength and low frequency limit, the bulk scalar channel of the quasinormal modes corresponds to the sound mode fluctuation of the fluid on the AdS boundary~\cite{Kovtun:2005ev}. While in the present case, the bulk gravitational background is the first order perturbed geometry with hydrodynamic fluctuation eq.(\ref{b4}), then the quasinormal mode will be obtained by further perturb the perturbed geometry eq.(\ref{b4}). However, the quasinormal mode perturbation is different with the hydrodynamic fluctuation from the bulk in Section.\ref{subsec:hydroperturb}, in which the quasinormal mode perturbation (the Lie derivative on metric) can be viewed as the probe field propagating in the unperturbed geometry, while the bulk hydrodynamic perturbation (the partial or covariant derivative on hydrodynamic variables) will cause the original geometry be corrected by higher derivative terms.

To determine the bulk scalar quasinormal mode dual to the phonon (sound normal mode) scattering in the acoustic black hole eq.(\ref{accut3}) formed in the fluid at cutoff surface $r=r_{\rm c}$, note that the normal mode perturbation of the fluid at cutoff surface are listed in eq.(\ref{normmode}), where $\delta\epsilon_{\rm c}$ and $\delta p_{\rm c}$ are acting on $r_{\rm h}$, i.e. the temperature of the fluid. From the bulk quasinormal mode perturbation side, the corresponding metric perturbations at $r=r_{\rm c}$ and at the acoustic horizon $(t=0,\cdots,0,z=z_{\rm sh})$, can be obtained from normal mode variation on the perturbed geometry eq.(\ref{b4}), which are
\be\label{normmodebulk}
\delta g_{rt}&=&\frac{\delta u_t(z)|_{z_{\rm sh}}}{f(r_{\rm c})}+\frac{du_t r_{\rm h}^{d-1}}{2r_{\rm c}^df(r_{\rm c})^{3/2}}\delta r_{\rm h},\quad \delta g_{rz}=\frac{\delta u_z(z)|_{z_{\rm sh}}}{f(r_{\rm c})}+\frac{du_z r_{\rm h}^{d-1}}{2r_{\rm c}^d f(r_{\rm c})^{3/2}}\delta r_{\rm h},\nno\\
\delta g_{tt}&=& \left(\theta u_t^2\mathfrak{s}'_1(r_{\rm c})+\frac{\theta}{d-1}P_{tt}\mathfrak{s}'_2(r_{\rm c})
+2a_t u_t\mathfrak{v}'(r_{\rm c})+\sigma_{tt}\mathfrak{t}'(r_{\rm c})\right)\delta r_{\rm h},\nno\\
\delta g_{tz}&=&\delta g_{zt}=\left(\theta u_t u_z\mathfrak{s}'_1(r_{\rm c})+\frac{\theta}{d-1}u_t u_z\mathfrak{s}'_2(r_{\rm c})
+(a_t u_z+a_z u_t)\mathfrak{v}'(r_{\rm c})+\sigma_{tz}\mathfrak{t}'(r_{\rm c})\right)\delta r_{\rm h},\nno\\
\delta g_{zz}&=& \left(\theta u_z^2\mathfrak{s}'_1(r_{\rm c})+\frac{\theta}{d-1}P_{zz}\mathfrak{s}'_2(r_{\rm c})
+2a_z u_z\mathfrak{v}'(r_{\rm c})+\sigma_{zz}\mathfrak{t}'(r_{\rm c})\right)\delta r_{\rm h},\nno\\
\delta g_{ab}&=&\left(\mathfrak{s}'_2(r_{\rm c})-\mathfrak{t}'(r_{\rm c})\right)\frac{\theta}{d-1}\delta_{ab}\delta r_{\rm h}, \quad ({\rm with}\quad a,b\neq t,z),
\ee
where $'$ is the partial derivative with respect to $r_{\rm h}$, and $\delta r_{\rm h}=\delta r_{\rm h}(r_{\rm c},t,z)$. The above variables can be combined into
\be\label{scalarmode}
Z_{\rm c}&=& u_z^2\delta g_{tt}+u_t^2\delta g_{zz}-2u_t u_z\delta g_{tz}+\sum_a\delta g_{aa}=\theta\mathfrak{s}'_2(r_{\rm c})\delta r_{\rm h},\nno\\
{\rm or},\quad Z_{\rm c}&=&-\frac{2r_{\rm c}^df(r_{\rm c})^{3/2}\theta\mathfrak{s}'_2(r_{\rm c})}{dr_{\rm h}^{d-1}}\left(u^t\delta g_{rt}+u^z\delta g_{rz}\right),
\ee
which form a gauge invariant scalar field $Z_{\rm c}=Z_{\rm c}(r_{\rm c},t,z)$-the sound channel of the bulk quasinormal mode (thanks to the remaining $SO(d-2)$ rotational symmetry of the background spacetime).

Together with normal mode perturbation of the longitudinal part of eq.(\ref{TaTb}), when omitting the subleading second order derivative terms, it is
\be\label{normmodefluid}
\left(\partial_\mu\ln r_{\rm h}+\frac{\partial_\mu}{d-1}\right)\delta u^\mu(t,z)=-u^\mu \partial_\mu\left(\frac{\delta r_{\rm h}(r_{\rm c},t,z) }{r_{\rm h}}\right)
\ee
and variation of eq.(\ref{vpotential}), the phonon field $\delta\psi(t,z)$ is
\be
\delta u^\mu(t,z)=\frac{P^{\nu\mu} \partial_\nu \delta\psi(t,z)}{\sqrt{-\partial_\al\psi \partial^\al\psi}}
\ee
for $\mu=t,z$. Then the one-to-one map between $Z_{\rm c}$ and the phonon $\delta\psi(t,z)$ is
\be\label{normmodefluid2}
\left(\partial_\mu\ln r_{\rm h}+\frac{\partial_\mu}{d-1}\right)\frac{P^{\nu\mu} \partial_\nu \delta\psi(t,z)}{\sqrt{-\partial_\al\psi \partial^\al\psi}}=-u^\mu \partial_\mu\left(\frac{Z_{\rm c}(r_{\rm c},t,z) }{\theta\mathfrak{s}'_2(r_{\rm c})r_{\rm h}}\right).
\ee

Furthermore, the dual operator of $Z_{\rm c}$ on the asymptotical boundary $r_{\rm c}\rightarrow \infty$ is also a scalar operator $\mathcal{\hat{O}}$ combined from the components of the boundary stress tensor $T^{tt}$,  $T^{t z}$,  $T^{zz}$ and $T_{a}^{a}$. In addition, the boundary stress tensor couples to the source part of the boundary metric via
\be \int_{\partial \mathcal{M}}T^{\mu\nu}\gamma_{\mu\nu}^{(0)}\sim \int _{\partial \mathcal{M}}\mathcal{\hat{O}}Z_0.
\ee
where $Z_0$ is the source term of $Z_{\rm c}$ and $\gamma_{\mu\nu}^{(0)}$ is the source term of the induced metric on the AdS boundary, respectively. Therefore, there is a duality between the sound channel quasinormal mode $Z_{\rm c}$ propagating in the bulk perturbed AdS black brane and the phonon $\delta\psi$ scattering in the acoustic black hole geometry formed from the fluid on the boundary or cutoff surface.

%%%%%%%%%%%%%%%%%%%%%%%%%%%%%%%%%%%%%%%%%%%%%%%%%%%%%%%%%%%%%%%%%%%%%%
\section{Conclusions and discussions}\label{sec:conclusion}
In this paper, we realized the holographic description of the acoustic black hole based on the formalism of the fluid/gravity correspondence. An acoustic black hole geometry formed in the fluid at finite timelike cutoff surface (membrane) in a neutral boosted black brane in asymptotically AdS spacetime was constructed, based on the matching between the conservation equation of fluid stress tensor and the constraint equations of the bulk Einstein equation. Besides, it was showed that the bulk dual of the acoustic black hole is the AdS black brane corrected with first order hydrodynamic fluctuation. Moreover, we determined the connection between the temperature of the acoustic black hole and the Hawking temperature of the real AdS black brane in the bulk. What's more, we showed that, the phonon field, which comes from the normal mode excitation of the fluid at the cutoff surface and scatters in the acoustic black hole geometry, is dual to the scalar field-the sound channel of quasinormal modes propagating in the bulk perturbed AdS black brane. Therefore, we pointed out that, from the viewpoint of the fluid/gravity duality, the acoustic black hole formed in the fluid is no longer just an analogous model of the real black hole. The remarkable connection between the two seemly different systems sheds lights on the study of the analogous gravitational models, which aim to acquire insights for studying various phenomena in the presence of gravity, such as the Hawking radiation. According to our results, the appearance of the acoustic black hole in the fluid located on the finite timelike cutoff surface indeed corresponds to a modification or perturbation to the geometry of the bulk AdS black brane, and the detecting of the Hawking-like temperature of the acoustic horizon can indeed give us some information about the Hawking temperature of the real black brane, at least in the asymptotically AdS or Lifshitz spacetime cases. \omits{Furthermore, the suggestions of the Hawking radiation is merely a kinematic effect need to be reconsidered due to the correspondence between the dynamics governing the fluid and the gravity \cite{Visser:1997yu,Huang:2007tw}.} There are many interesting related problems to explore such as more on the duality between the acoustic black hole and its dual bulk AdS black brane, e.g., comparing the scattering of phonons by the acoustic black hole and the same process of sound channel quasinormal mode in the bulk AdS black brane, studying the acoustic black hole in the fluid with anomalies, more about the experimentally testable effect of the temperature of the acoustic black hole on its dual bulk black hole, the supersonic phenomena in the Quark-Gluon-Plasma, finding the effective action (such as in \cite{Kovtun:2014hpa}) to describe the holographic acoustic black holes, analyzing acoustic black holes in many other condensed matter systems etc.

%%%%%%%%%%%%%%%%%%%%%%%%%%%%%%%%%%%%%%%%%%%%%%%%%%%%%%%%%%%%%%%%%%%%%%
%\section{Absorption of phonon from the acoustic black hole}
%%%%%%%%%%%%%%%%%%%%%%%%%%%%%%%%%%%%%%%%%%%%%%%%%%%%%%%%%%%%%%%%%%%%%%
 %Now let us study the absorption of phonons by the acoustic black hole from cutoff surfaces.

%%%%%%%%%%%%%%%%%%%%%%%%%%%%%%%%%%%%%%%%%%%%%%%%%%%%%%%%%%%%%%%%%%%%%%
%\section{Holographic Wilsonian RG flow}
%%%%%%%%%%%%%%%%%%%%%%%%%%%%%%%%%%%%%%%%%%%%%%%%%%%%%%%%%%%%%%%%%%%%%%

%%%%%%%%%%%%%%%%%%%%%%%%%%%%%%%%%%%%%%%%%%%%%%%%%%%%%%%%%%%%%%%%%%%%%%
%\subsection{Near horizon limit}
%%%%%%%%%%%%%%%%%%%%%%%%%%%%%%%%%%%%%%%%%%%%%%%%%%%%%%%%%%%%%%%%%%%%%%

%%%%%%%%%%%%%%%%%%%%%%%%%%%%%%%%%%%%%%%%%%%%%%%%%%%%%%%%%%%%%%%%%%%%%%
%\subsection{Asymptotic limit}
%%%%%%%%%%%%%%%%%%%%%%%%%%%%%%%%%%%%%%%%%%%%%%%%%%%%%%%%%%%%%%%%%%%%%%

%%%%%%%%%%%%%%%%%%%%%%%%%%%%%%%%%%%%%%%%%%%%%%%%%%%%%%%%%%%%%%%%%%%%%%
%\section{Beyond the Linearized Perturbation}
%%%%%%%%%%%%%%%%%%%%%%%%%%%%%%%%%%%%%%%%%%%%%%%%%%%%%%%%%%%%%%%%%%%%%%

%%%%%%%%%%%%%%%%%%%%%%%%%%%%%%%%%%%%%%%%%%%%%%%%%%%%%%%%%%%%%%%%%%%%%%
\section*{Acknowledgement}
%%%%%%%%%%%%%%%%%%%%%%%%%%%%%%%%%%%%%%%%%%%%%%%%%%%%%%%%%%%%%%%%%%%%%%
We would like to thank Rong-Gen Cai, Giuseppe Policastro and Hong-Bao Zhang for useful discussions. X.H.G. was supported by the NSFC (No.~11375110). J.R.S. was supported by the National Science Foundation of China under Grant No.~11205058 and the Open Project Program of State Key Laboratory of Theoretical Physics, Institute of Theoretical Physics, Chinese Academy of Sciences, China (No.~Y5KF161CJ1). Y.T. was partially supported by NSFC with Grant No.~11475179. X.N.W. was was partially supported by NSFC with Grant No.~11175245. Y.L.Z. thanks the support from CASTS (No.~104R891003) at NTU, MOST grant (No.~104-2811-M-002-080) and the support from SKLTP (No.~09KL141Y31) during the KITPC workshop.

%%%%%%%%%%%%%%%%%%%%%%%%%%%%%%%%%%%%%%%%%%%%%%%%%%%%%%%%%%%%%%%%%%%%%%
\begin{appendix}
%%%%%%%%%%%%%%%%%%%%%%%%%%%%%%%%%%%%%%%%%%%%%%%%%%%%%%%%%%%%%%%%%%%%%%
%%%%%%%%%%%%%%%%%%%%%%%%%%%%%%%%%%%%%%%%%%%%%%%%%%%%%%%%%%%%%%%%%%%%%%
\section{First order metric corrections}\label{sec:1st correc}
%%%%%%%%%%%%%%%%%%%%%%%%%%%%%%%%%%%%%%%%%%%%%%%%%%%%%%%%%%%%%%%%%%%%%%
The non-vanishing components of the first order metric corrections in eq.(\ref{mecor1st}) are
\be
g^{(1)}_{tt}(r,z)&=& \left(\theta u_t^2 \mathfrak{s}_1(r)+\frac{\theta}{d-1}P_{tt}\mathfrak{s}_2(r)+2a_t u_t \mathfrak{v}_2(r)+\left(u_t a_t-\frac{\theta}{d-1}P_{tt}\right)\mathfrak{t}(r)\right),\nno\\
g^{(1)}_{tz}(r,z)&=& g^{(1)}_{zt}(r,z)=\bigg(\theta u_t u_z \mathfrak{s}_1(r)+\frac{\theta}{d-1}P_{tz}\mathfrak{s}_2(r)+\left(a_t u_z +a_z u_t \right)\mathfrak{v}_2(r)\nno\\&&+\left(\frac{1}{2}\left(u_z a_t+\partial_z u_t+u_t a_z\right)-\frac{\theta}{d-1}P_{tz}\right)\mathfrak{t}(r)\bigg),\nno\\
g^{(1)}_{zz}(r,z)&=& \left(\theta u_z^2 \mathfrak{s}_1(r)+\frac{\theta}{d-1}P_{zz}\mathfrak{s}_2(r)+2a_z u_z \mathfrak{v}_2(r)+\left(u_z a_z+\partial_z u_z-\frac{\theta}{d-1}P_{zz}\right)\mathfrak{t}(r)\right),\nno\\
g^{(1)}_{ab}(r,z)&=&\bigg(\mathfrak{s}_2(r)-\mathfrak{t}(r)\bigg)\frac{\theta}{d-1}\delta_{ab} \quad ({\rm with}\quad a,b\neq t,z),
\ee
where at the acoustic horizon $z=z_{\rm sh}$, we have
\be
u_t &=& \frac{1}{\al_{\rm c}\sqrt{f(r_{\rm c})}},\quad u_z=\frac{\hat{c}_{\rm s}}{\al_{\rm c}},\nno\\
\theta &=& \partial_z u^z= -\frac{2\pi T_{\rm sh}}{\al_{\rm c}\sqrt{f(r_{\rm c})}}+\frac{d\hat{c}_{\rm s}r_{\rm h}^{d-1}}{2\al_{\rm c}f(r_{\rm c})r^d}\partial_z r_{\rm h}|_{z_{\rm sh}},\nno\\
a_t &=& u^z \partial_z u_t= -\frac{2\pi\hat{c}_{\rm s}^2 T_{\rm sh}}{\al_{\rm c}^2}+\frac{d\hat{c}_{\rm s}^3 r_{\rm h}^{d-1}}{2\al_{\rm c}^2\sqrt{f(r_{\rm c})}r^d}\partial_z r_{\rm h}|_{z_{\rm sh}},\nno\\
a_z &=& u^z \partial_z u_z= -\frac{2\pi \hat{c}_{\rm s} T_{\rm sh}}{\al_{\rm c}^2\sqrt{f(r_{\rm c})}}+\frac{d\hat{c}_{\rm s}^2 r_{\rm h}^{d-1}}{2\al_{\rm c}^2 f(r_{\rm c})r^d}\partial_z r_{\rm h}|_{z_{\rm sh}}.
\ee

%%%%%%%%%%%%%%%%%%%%%%%%%%%%%%%%%%%%%%%%%%%%%%%%%%%%%%%%%%%%%%%%%%%%%%
\section{3-dimensional relativistic rotating acoustic black hole}\label{sec:3d acoustic}
%%%%%%%%%%%%%%%%%%%%%%%%%%%%%%%%%%%%%%%%%%%%%%%%%%%%%%%%%%%%%%%%%%%%%%
For simplicity we focus on 3-dimensional irrotational fluid and adopting the polar coordinates, then eq.(\ref{accut}) is written as (omitting the conformal factor)
\be\label{accut3d}ds_{\rm ac}^2&\sim&
-\left(\hat{c}_{\rm s}^2f(r_{\rm c})-\alpha_{\rm c}^2 \bar{\tilde{u}}_{i}^2\right)dt^2+2\al_{\rm c}^2\bar{\tilde{u}}_{0}\left(\bar{\tilde{u}}_{\varrho}d\varrho
+\bar{\tilde{u}}_{\varphi}d\varphi\right)dt+\al_{\rm c}^2\left(\tilde{u}_{0\varrho}d\varrho
+\tilde{u}_{0\varphi}d\varphi\right)^2\nno\\
&&+\varrho^2d\varphi^2+d\varrho^2,\ee
with $\bar{\tilde{u}}_{\mu}=\sqrt{f(r_{\rm c})}\gamma(1, -\frac{a}{\varrho}, b)$, where $a$ and $b$ are constants. Eq.(\ref{accut3d}) is just the relativistic counterpart of the 3-dimensional rotational acoustic black hole formed in the draining bathtub. When putting the cutoff surface to the AdS boundary and taking the non-relativistic limit, eq.(\ref{accut3d}) reduces to the known result
\be
\label{accut3dnr}ds_{\rm ac}^2&\sim&
-\frac 1 2\left(1-\frac{a^2+b^2}{\varrho^2}\right)dt^2 -\frac{a}{\varrho}d\varrho dt
+b d\varphi dt +\varrho^2d\varphi^2+d\varrho^2\ee
that was studied in \cite{Visser:1997ux}.

\end{appendix}
%%%%%%%%%%%%%%%%%%%%%%%%%%%%%%%%%%%%%%%%%%%%%%%%%%%%%%%%%%%%%%%%%%%%%%

%%%%%%%%%%%%%%%%%%%%%%%%%%%%%%%%%%%%%%%%%%%%%%%%%%%%%%%%%%%%%%%%%%%%%%

\end{document}